\documentclass[conference, 10pt]{IEEEtran}

\usepackage{cite}
\usepackage{orcidlink}
\usepackage{amsmath}
\usepackage{url}
\usepackage{graphicx}
\usepackage{color}
\usepackage{placeins}
\usepackage{float}
\usepackage{tabularx,colortbl}
\usepackage{ifthen}

\makeatletter

\newcounter{author}
\renewcommand{\author}[2][]{
   \stepcounter{author}
   \@namedef{author@\theauthor}{#2}
   \@namedef{authorlabel@\theauthor}{#1}
}

\newcounter{address}
\newcommand{\address}[2][]{
   \stepcounter{address}
   \@namedef{address@\theaddress}{#2}
   \@namedef{addresslabel@\theaddress}{#1}
}

\newcommand{\alsep}{and}

\def\newmaketitle{\par%
  \begingroup%
  \normalfont%
  \def\thefootnote{}
  \def\footnotemark{}
  \let\@makefnmark\relax
  \footnotesize
  \footnotesep 0.7\baselineskip
  \normalsize%
  \twocolumn[\thenewmaketitle\@IEEEaftertitletext]%
  \if@IEEEusingpubid
     \enlargethispage{-\@IEEEpubidpullup}%
  \fi
  \endgroup
  \setcounter{footnote}{0}\let\maketitle\relax\let\@maketitle\relax
  \gdef\@thanks{}%
  \let\thanks\relax}

\def\thenewmaketitle{
  \newpage
  \begin{center}%
    \vskip0.2em{\Huge\@IEEEcompsoconly{\sffamily}\@IEEEcompsocconfonly{\normalfont\normalsize\vskip 2\@IEEEnormalsizeunitybaselineskip
   \bfseries\large}\@title\par}\vskip1.0em\par%
    \vspace{1ex}
    \newcounter{c@author}
    \newcounter{c@tmp}
    \ifthenelse{\value{author}=2}{%
      \newcommand{\liand}{ and }}{%
      \newcommand{\liand}{, and }}
    \ifthenelse{\value{address}<2}{%
      \@nameuse{author@1}%
      \stepcounter{c@author}%
      \whiledo{\value{c@author}<\value{author}}{%
        \setcounter{c@tmp}{\value{author}}%
        \addtocounter{c@tmp}{-\value{c@author}}%
        \ifthenelse{\value{c@tmp}=1}{%
          \renewcommand{\alsep}{\liand}}{\renewcommand{\alsep}{, }}%
        \stepcounter{c@author}\alsep \@nameuse{author@\thec@author}}\\%
    }
    {
      \@nameuse{author@1}${}^{(\ref{\@nameuse{authorlabel@1}})}$%
      \stepcounter{c@author}%
      \whiledo{\value{c@author}<\value{author}}{%
      \setcounter{c@tmp}{\value{author}}%
      \addtocounter{c@tmp}{-\value{c@author}}%
      \ifthenelse{\value{c@tmp}=1}{%
        \renewcommand{\alsep}{\liand}}{\renewcommand{\alsep}{, }}%
      \stepcounter{c@author}\alsep \@nameuse{author@\thec@author}%
        ${}^{(\ref{\@nameuse{authorlabel@\thec@author}})}$%
      }
    }
    \vspace{0.2ex}

    \ifthenelse{\value{address}>0}{%
      \ifthenelse{\value{address}=1}{
        {\@nameuse{address@1}}
      }
      {
        \newcounter{c@address}

        \begin{center}
        \whiledo{\value{c@address}<\value{address}}
        {
          \refstepcounter{c@address}
            ${}^{(\thec@address)}$\,%
              \label{\@nameuse{addresslabel@\thec@address}}%
              \@nameuse{address@\thec@address}\\ %
        }
        \end{center}
      } 
    }
    {
      \relax
    }
  \end{center}
}

\makeatother

\title{A Lumped-Element Electrical Model of the Human Head for Brain-Oriented Applications}

\author[org1]{Angelo~Faccia}
\author[org1]{Ermanno~Citraro}
\author[org1]{Francesco~P.~Andriulli}

\address[org1]{Department of Electronics and Telecommunications, Politecnico di Torino, 10129 Turin, Italy}

\begin{document}

\newmaketitle

\begin{abstract}
In this work, we present a compact surrogate circuit for electro-quasi-static (EQS) head modeling. A three-shell geometry (brain, skull, scalp) is considered, and each layer is modeled through radial and tangential pathways, implemented as RC branches.
Frequency-dependent tissue conductivity and permittivity are mapped into dispersive resistive and capacitive elements. The model is validated against a semi-analytical spherical-harmonics reference solution over multiple geometrical configurations and operating frequencies, demonstrating good agreement.
Neglecting dispersion and capacitive pathways can lead to an overestimation of scalp potentials over the considered frequency range, highlighting the need for dispersive RC circuit modeling.
\end{abstract}

\section{Introduction}
Accurate yet computationally efficient models for estimating electric potential and current density distributions in head tissues are essential for the design, safety assessment, and optimization of neuro-sensing and neuro-stimulation electronic systems.

Many technologies of practical interest operate at sub-MHz frequencies, where electro-quasi-static (EQS) assumptions are typically adopted~\cite{plonsey1967considerations}. This range spans applications from EEG source imaging~\cite{Hallez2007,grech2008review} to stimulation modalities, including conventional deep brain stimulation (DBS) and emerging ultrahigh-frequency DBS protocols~\cite{whitmer2012high,harmsen2019p}.

Numerical techniques (e.g., FEM/FDM/BEM)~\cite{jin2015theory} can incorporate realistic anatomy and tissue heterogeneity, but their computational cost and limited compatibility with circuit simulators may hinder rapid prototyping of electronic systems and real-time closed-loop evaluation. To mitigate this, circuit-oriented alternatives such as resistor mesh models have been proposed~\cite{franceries2003solution}. However, these approaches are often purely resistive and therefore neglect capacitive and dispersive effects, which can become increasingly relevant as frequency rises~\cite{Gaugain2023}.

To address these limitations, this work introduces a compact lumped RC equivalent circuit of the head derived from a canonical three-layer spherical geometry (brain, skull, scalp). The proposed network reproduces the peak scalp potential generated by an intracranial dipolar source up to 50~kHz, explicitly accounting for frequency-dependent tissue conductivity and permittivity as well as displacement-current pathways via capacitive branches.
The model is designed to support both sensing and stimulation applications via the reciprocity theorem~\cite{rush2008eeg}.

The proposed surrogate circuit is validated against a semi-analytical reference based on scalar spherical harmonics (SSH), while quantifying the error incurred when dispersion and capacitive pathways are omitted.

\section{Background and Notation}
In the EQS regime, inductive effects are negligible and the electric field is irrotational, $\mathbf{E}=-\nabla V$, \cite{plonsey1967considerations}. Under time-harmonic conditions, tissue electrical behavior is captured through complex, dispersive conductivity
\begin{equation}
\sigma_c(\omega)=\sigma(\omega)+j\omega\varepsilon(\omega),
\end{equation}
which combines conductive and capacitive responses. The governing equation for the scalar potential reads
\begin{equation}
\nabla\cdot\left(\sigma_c(\omega)\nabla V\right)=\nabla\cdot \mathbf{J}_i,
\end{equation}
where $\mathbf{J}_i$ represents impressed source currents. At tissue interfaces, continuity of the electric potential and of the normal component of the total current density yields
\begin{equation}
V_- = V_+, \qquad
\sigma_{c,-}\, E^n_-
=
\sigma_{c,+}\, E^n_+,
\label{eq:bound}
\end{equation}
where $E^n = \hat{\mathbf{n}} \cdot \mathbf{E} = -\frac{\partial V}{\partial n}$ is the normal component of the electric field at the interface, with $\hat{\mathbf{n}}$ pointing outward from inner domain $-$ to outer domain $+$.

\begin{figure}
    \centering
    \includegraphics[width=1\linewidth]{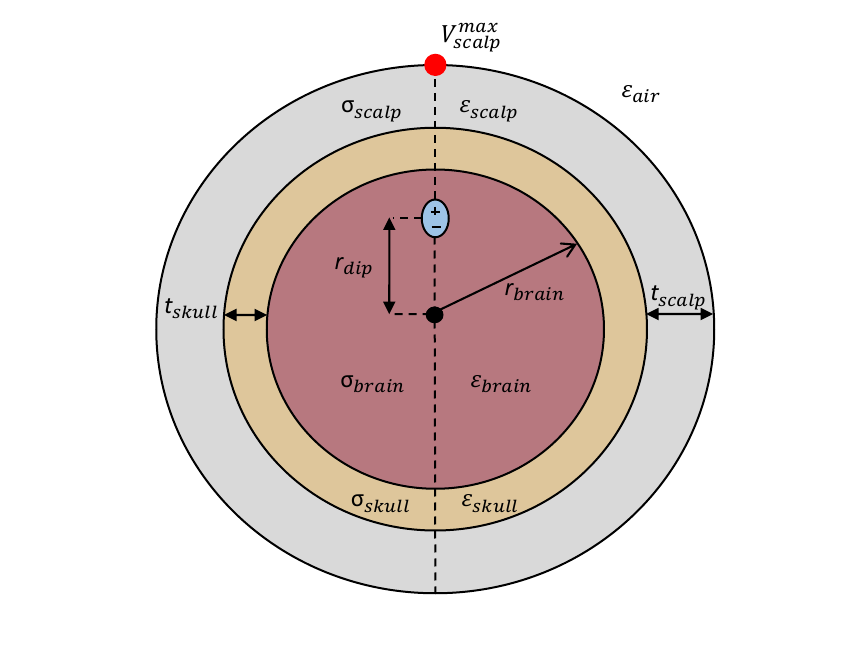}
    \caption{
Three-layer spherical head model. 
A radial current dipole is located within the brain at a distance \(r_{\mathrm{dip}}\) from the center. 
Each tissue layer is characterized by frequency-dependent conductivity \(\sigma(\omega)\) and permittivity \(\varepsilon(\omega)\). 
Standard geometrical parameters are set as: brain radius \(r_{\mathrm{brain}} = 7.91~\mathrm{cm}\), skull thickness \(t_{\mathrm{skull}} = 5.9~\mathrm{mm}\), and scalp thickness \(t_{\mathrm{scalp}} = 7~\mathrm{mm}\).
}
    \label{fig:headmodel}
\end{figure}

We consider a piecewise-homogeneous, concentric, three-shell spherical head model (brain, skull, scalp) embedded in an infinite air region, as in Fig.~\ref{fig:headmodel}. A radially oriented current dipole is placed in the brain at radial distance $r_{\mathrm{dip}}$ from the center with dipole moment \(p_r\). 

Let \(\sigma^c_i\) denote the complex conductivity of the \(i\)-th layer, and define the geometrical ratios as \(\eta = \frac{r_{\text{dip}}}{r_1}\) and \(\psi_{ij} = \frac{r_i}{r_j}\), where \(r_i\) represents the radius of the interface between layers \(i\) and \(i+1\). In this geometry, the peak scalp potential can be computed via SSH expansion~\cite{arthur1970effect} as
\begin{equation}
V^{\mathrm{SSH}}_{(\omega,r_3)} = \frac{p_r}{4\pi r_{3}^2} 
\displaystyle\sum_{l=1}^{\infty} A_{4}(l,\omega),
\label{eq:anal}
\end{equation}
\begin{equation}
A_{4}(l,\omega)=\frac{l(2l+1)^3\sigma^c_{2}\sigma^c_{3}\eta^{l-1}\psi_{13}^{l-1}}
{l(l+1)(\psi_{23}^{2l+1}X_{1} + \psi_{13}^{2l+1}X_{2} + \psi_{12}^{2l+1}X_{3}) + \tilde{X}},
\end{equation}
where layers are indexed as \(1\) for the brain, \(2\) for the skull, \(3\) for the scalp, and \(4\) for the surrounding air. The auxiliary terms are defined as
\[
X_1 = \tilde{\sigma}^c_{21} \sigma^c_{32} \sigma^c_{43}, \quad
X_2 = \sigma^c_{21} \tilde{\sigma}^c_{23} \sigma^c_{43}, \quad
X_3 = \sigma^c_{21} \sigma^c_{32} \tilde{\sigma}^c_{43},
\]
\[
\tilde{X} = \tilde{\sigma}^c_{21} \tilde{\sigma}^c_{32} \tilde{\sigma}^c_{43},
\]
with \(\sigma^c_{ij} = \sigma^c_i - \sigma^c_j\) and \(\tilde{\sigma}^c_{ij} = (l+1)\sigma^c_i + l\sigma^c_j\).

This semi-analytical solution provides a rigorous benchmark, but it is inherently tied to spherical geometries and does not directly translate into a circuit netlist. These limitations motivate the design of a compact circuit surrogate that preserves the relevant physics in the same frequency band.
\begin{figure*}
    \centering
    \includegraphics[width=1\linewidth]{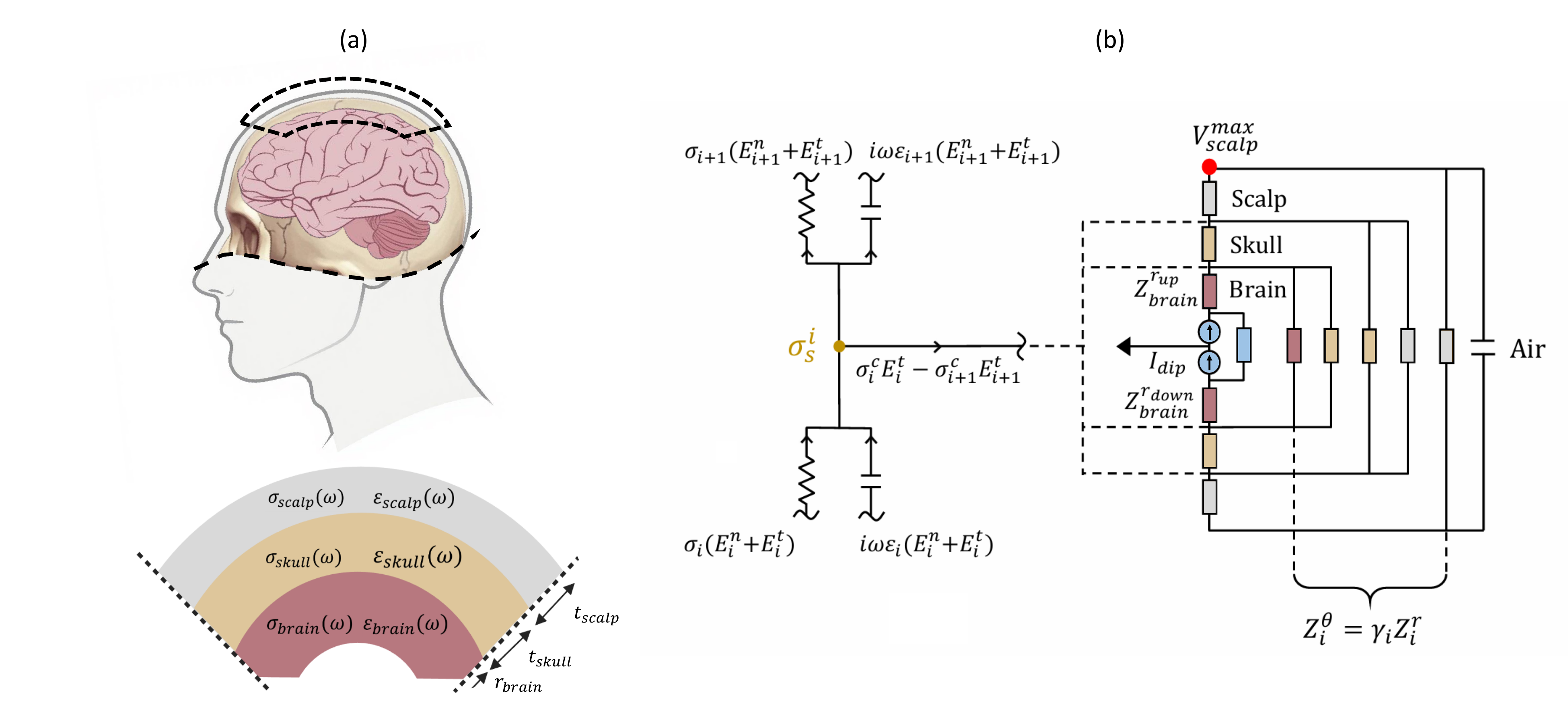}
    \caption{(a) Schematic representation of a piecewise homogeneous three layer spherical head approximation (b) Proposed lumped-circuit model with a radially oriented dipolar brain source. Vertically aligned branches implement the radial impedances $Z_i^{r}$ of each tissue layer; horizontally aligned branches implement the tangential impedances $Z_i^{\theta}$. Kirchhoff's current law (KCL) at every node enforces the Neumann boundary condition between adjacent layers $i$ and $i+1$, including the effect of the interfacial surface charge density $\sigma^i_{s}$ that arises from permittivity discontinuities. The brain radial impedances $Z_{\text{brain}}^{r_{\mathrm{up}}}$ and $Z_{\text{brain}}^{r_{\mathrm{down}}}$ differ according to dipole eccentricity. The point $V_{\text{scalp}}^{\max}$ denotes the maximum scalp voltage. Image designed with Freepik.
}
    \label{fig:headmodel2}
\end{figure*}
\section{Circuit Modeling Procedure}
\subsection{Circuit Topology}
Fig.~\ref{fig:headmodel2}(b) illustrates the proposed circuit topology.

A naive series stacking of layer impedances along the radial direction cannot reproduce the current partitioning produced by an internal dipole. Part of the current must cross successive shells, while another part flows within each shell before returning. To capture these mechanisms with minimal complexity, each tissue layer $i$ is represented by:
\begin{itemize}
\item a \emph{radial} impedance branch $Z_i^{r}$ modeling through-layer current flow;
\item a \emph{tangential} impedance branch $Z_i^{\theta}$ modeling intra-layer return currents.
\end{itemize}
As shown in Fig.~\ref{fig:headmodel2}(b), Kirchhoff's current law at radial nodes mirrors the Neumann boundary condition on total current density (\ref{eq:bound}), ensuring physically consistent current continuity. In this representation, interfacial surface charge accumulation is naturally associated with the capacitive behavior at radial nodes.

\subsection{Circuit Parametrization}
Each branch is implemented as a parallel RC element. Resistances scale inversely with conductivity, and capacitances scale with permittivity, while geometry is embedded through layer-dependent factors $\Gamma_i$:
\begin{equation}
R_i = \frac{\Gamma_i{(r_i,t_i)}}{\sigma_i(\omega)},\qquad C_i = \frac{\varepsilon_i(\omega)}{\Gamma_i{(r_i,t_i)}},
\end{equation}
so that the frequency dependence of $\sigma(\omega)$ and $\varepsilon(\omega)$ directly induces dispersive resistances $R_i(\omega)$ and capacitances $C_i(\omega)$.

A dimensionless coefficient is introduced to couple tangential and radial pathways,
\begin{equation}
\gamma_i=\frac{Z_i^{\theta}}{Z_i^{r}}=\frac{\Gamma_i^{\theta}}{\Gamma_i^{r}},
\end{equation}
which depends only on shell radii and thicknesses and provides a compact control of current partitioning between radial and tangential branches.
Inter-subject morphological variability is captured while preserving model compactness by parameterizing $\gamma_i$ as first and second order polynomial functions of radius ratios $\psi_{i,j}=r_i/r_j$. 

The mapping $\gamma_i(\psi)$ is identified by matching the circuit response to the analytical SSH reference over a sweep of geometries with non-dispersive, static tissue properties, so that the optimization targets geometry dependence only and the frequency-dispersive regime can later be used to independently validate the circuit model.

Finally, global head-size changes are accounted for by uniformly scaling the impedances proportionally to $1/r_{\mathrm{scalp}}^2$.
\subsection{Source Modeling}
The neural source is modeled as a current source $I_{\mathrm{dip}}$ related to the dipole moment $p_r$ through an effective dipole length $d$:
\begin{equation}
I_{\mathrm{dip}}=\frac{p_r}{d}.
\end{equation}
When the dipole is off-center, current paths in the brain become asymmetric. This is captured by splitting the brain radial impedance into two branches, $Z_{\mathrm{brain}}^{r_{\mathrm{up}}}$ and $Z_{\mathrm{brain}}^{r_{\mathrm{down}}}$, redistributed via an eccentricity-dependent asymmetry parameter $\alpha(\eta)$, where $\eta=r_{\mathrm{dip}}/r_1$ and $r_1$ is the brain radius:
\begin{equation}
\begin{aligned}
Z_{\mathrm{brain}}^{r_{\mathrm{up}}}&=Z_{\mathrm{brain}}^{r,0}\left[1-\alpha(\eta)\right],\\[4pt]
Z_{\mathrm{brain}}^{r_{\mathrm{down}}}&=Z_{\mathrm{brain}}^{r,0}\left[1+\alpha(\eta)\right].
\end{aligned}    
\end{equation}
The function $\alpha(\eta)$ is obtained by fitting optimized values to the SSH reference over a physiologically relevant eccentricity range of up to $\eta=0.965$, under the same fitting setup used for $\gamma_i(\psi)$, enabling accurate modeling of cortical source locations.

\section{Validation and Numerical Results}
Building on the previously identified geometry-dependent circuit parameters, the model is validated in a more comprehensive scenario where frequency-dependent electrical parameters are combined with concurrent variations in geometry and dipole position. In this extended setting, dispersive tissue behavior is captured by using the frequency-dependent conductivity and permittivity reported by Wagner \textit{et al.}~\cite{wagner2014}, and the ability of the model to correctly generalize is assessed. In addition, the error incurred when displacement currents and dispersion are omitted is explicitly quantified to assess the impact of these modeling assumptions.
\subsection{Agreement with the Semi-Analytical Reference}
A radial dipole moment of $p_r=15~\mathrm{nA\cdot m}$ is considered with effective length $d=1~\mathrm{mm}$. Multiple dipole eccentricities are tested, $\eta = 0.233$, $0.465$, $0.814$, $0.935$, and $0.966$, corresponding to dipole radial positions $r_{\mathrm{dip}} = 1.84~\mathrm{cm}$, $3.68~\mathrm{cm}$, $6.44~\mathrm{cm}$, $7.40~\mathrm{cm}$, and $7.64~\mathrm{cm}$.

The skull thickness is swept from $4.6~\mathrm{mm}$ to $8.2~\mathrm{mm}$, while the brain and scalp radii are held constant.

\begin{figure}[t]
    \centering
    \includegraphics[width=1\linewidth]{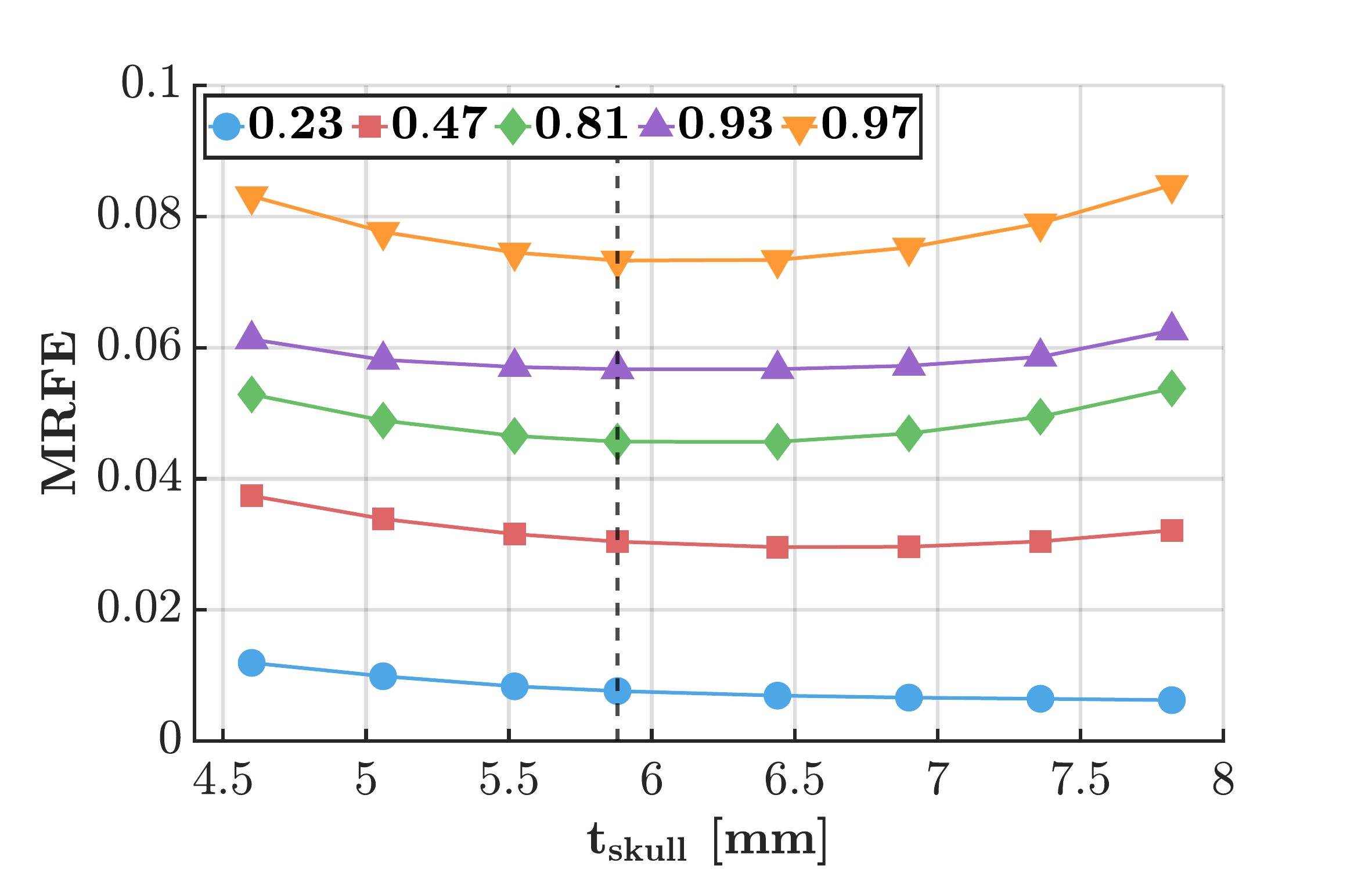}
    \caption{Mean Relative Frequency Error (MRFE) in \(|V_{\mathrm{scalp}}^{\mathrm{max}}|\) between the lumped model and the semi-analytical reference, evaluated over dipole eccentricity $\eta$ and skull thickness $t_{\mathrm{skull}}$. The dashed vertical line indicates the performances achieved with standard skull thickness.}
    \label{fig:mrfe_plot}
\end{figure}

To quantify the agreement with the SSH reference, we evaluate the mean relative frequency error (MRFE) for the peak scalp potential across all tested dipole eccentricities and skull thicknesses, using $N_f=75$ frequency points uniformly sampled between 10 Hz and 50 kHz. The resulting MRFE is reported in Fig.~\ref{fig:mrfe_plot}:
\begin{equation}
\mathrm{MRFE}=\frac{1}{N_f}\sum_{f}\left|\frac{V^{\mathrm{Circuit}}_{\mathrm{scalp},f}-V^{\mathrm{SSH}}_{\mathrm{scalp},f}}{V^{\mathrm{SSH}}_{\mathrm{scalp},f}}\right|.
\end{equation}

The MRFE increases moderately with dipole eccentricity and with shifts of the skull--scalp boundary toward either the brain or the outer air interface, with the highest sensitivity observed for the most eccentric dipoles. Overall, eccentricity has a stronger impact on the error than skull-thickness variations alone. Deviations become more noticeable at high eccentricity, where local geometric effects are harder to capture with a limited number of circuit elements; nevertheless, the MRFE values indicate good agreement with the semi-analytical reference.
\subsection{Impact of Dispersion and Displacement Currents}
Finally, we evaluate how capacitive and dispersive effects influence the amplitude of the circuit-predicted scalp potential.
As an illustrative example, we consider a highly eccentric dipole $(\eta = 0.935)$ located at radial coordinate $r_{\mathrm{dip}} = 7.40\,\mathrm{cm}$.

\begin{figure}[t]
    \centering
    \includegraphics[width=1\linewidth]{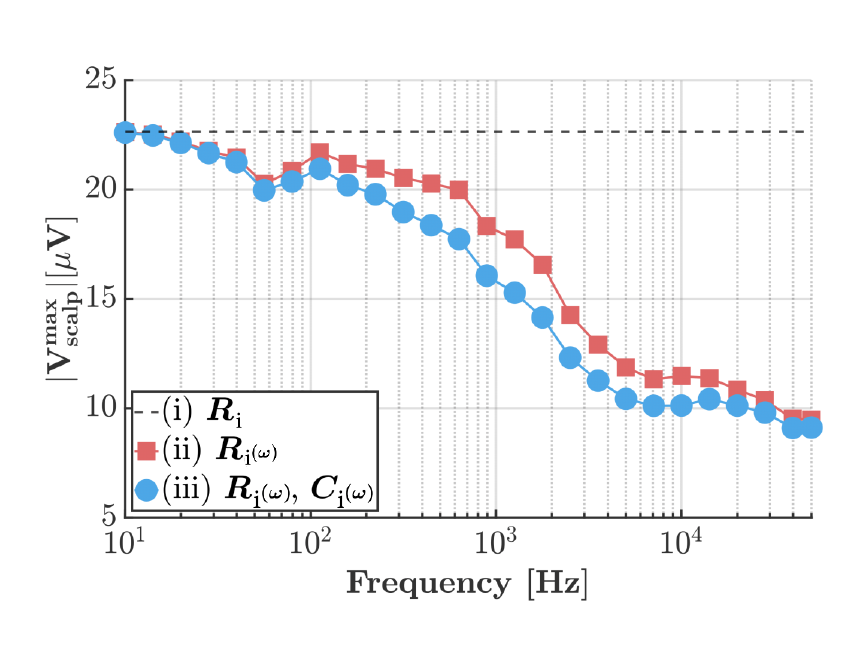}
    \caption{Frequency response of the maximum scalp potential magnitude $|V_{\mathrm{scalp}}^{\mathrm{max}}|$ predicted by the lumped circuit for a radial dipole at $r_{\mathrm{dip}}=7.40~\mathrm{cm}$ ($\eta=0.935$) in the three-shell head (cf. Fig.~\ref{fig:headmodel}). The three cases correspond to: (i) ohmic, non-dispersive $R_i$; (ii) dispersive $R_i(\omega)$ without displacement currents; and (iii) dispersive $R_i(\omega)$ and dispersive $C_i(\omega)$.}
    \label{fig:omitt}
\end{figure}

We compare three circuit configurations:
\begin{itemize}
\item (i) purely ohmic and non-dispersive: constant static resistances $R_i$ and no capacitances ($C_i=0$);
\item (ii) dispersive conductivity only: $R_i(\omega)$ with $C_i=0$;
\item (iii) dispersive conductivity and permittivity: $R_i(\sigma(\omega))$ and $C_i(\varepsilon(\omega))$.
\end{itemize}
As shown in Fig.~\ref{fig:omitt}, neglecting dispersion and displacement-current pathways can substantially overestimate the peak scalp voltage. The purely ohmic case yields the largest deviation, with errors becoming appreciable already at tens of hertz and exceeding $100\%$ toward the upper end of the investigated band. Including dispersive conductivity alone markedly reduces the error, but discrepancies above $10\%$ persist, peaking at $~17\%$ near 1.7 kHz. These findings indicate that accurate electrical head modeling, even below $50\,\mathrm{kHz}$, requires accounting for both frequency-dependent impedance and capacitive current pathways.
\section{Conclusion}
A compact lumped RC equivalent circuit of the human head was developed for EQS electrical modeling. Despite using a minimal set of circuit elements, the model shows good agreement with the semi-analytical reference while remaining compatible with standard circuit simulation environments. Incorporating frequency-dependent tissue properties and capacitive pathways is necessary to avoid systematic overestimation of scalp potentials when broadband behavior is considered. 

In conclusion, the proposed circuit provides a practical surrogate model for rapid device prototyping and circuit-level integration, and it can support simulations for adaptive neuromodulation. Future work will extend the framework to tangential dipoles, anisotropic conductivities, and more realistic head geometries while preserving circuit compactness.

\section*{ACKNOWLEDGMENT}
This work has received funding from the European Innovation Council (EIC) through the European Union's Horizon Europe research Programme under Grant 101046748 (Project CEREBRO) and from the European Research Council (ERC) through the HORIZON ERC Proof of Concept Grants under Grant 101189419 (Project TurboEEG).
\bibliographystyle{ieeetr}       
\bibliography{references}        

\end{document}